# Quantifying the dynamics of protein self-organization using deep learning analysis of atomic force microscopy data

Maxim Ziatdinov,[1] Shuai Zhang,[2,3] Orion Dollar[4], Jim Pfaendtner,[4] Chris Mundi,[3] Xin Li,[1] Harley Pyles,[5,6] David Baker,[5,6,7] James J. De Yoreo,[2,3] and Sergei V. Kalinin[1]

[1] Center for Nanophase Materials Sciences, Oak Ridge National Laboratory, Oak Ridge, TN, USA

[2] Materials Science and Engineering, University of Washington, Seattle, WA, USA

[3] Physical Sciences Division, Pacific Northwest National Laboratory, Richland, WA, USA

[4] Chemical Engineering, University of Washington, Seattle, WA, USA

[5] Department of Biochemistry, University of Washington, Seattle, WA, USA

[6] Institute for Protein Design, University of Washington, Seattle, WA, USA

[7] Howard Hughes Medical Institute, University of Washington, Seattle, WA, USA

Dynamics of protein self-assembly on inorganic surface and the resultant geometric patterns are visualized using high-speed atomic force microscopy. The time dynamics of the classical macroscopic descriptors such as 2D Fast Fourier Transforms (FFT), correlation and pair distribution function are explored using the unsupervised linear unmixing, demonstrating the presence of static ordered and dynamic disordered phases and establishing their time dynamics. The deep learning (DL)-based workflow is developed to analyze detailed particle dynamics on particle-by-particle level. Beyond the macroscopic descriptors, we utilize the knowledge of local particle geometries and configurations to explore the evolution of local geometries and reconstruct the interaction potential between the particles. Finally, we use the machine learning based feature extraction to define particle neighborhood free of physics constraints. This approach allowed separating the possible classes of particle behavior, identify the associated transition probabilities, and further extend this analysis to identify slow modes and associated configurations, allowing for systematic exploration and predictive modelling of the time dynamics of the system. Overall, this work establishes the DL based workflow for the analysis of the self-organization processes in complex systems from observational data and provides insight into the fundamental mechanisms.

In nature, protein self-assembly and directed morphogenesis of inorganic materials is an efficient bottom-up process of fabricating functional structures across scales with outstanding precision, efficiency, and adaptability.[1] Though attempts to engineer protein-protein and protein-inorganic interfaces to achieve a similar level of control and function in artificially designed systems have demonstrated some significant successes,[2-7] advancement is held back by an absence of design rules. Those rules must reflect the underlying energy landscape across which assembly and morphogenesis progress. Thus, probing the dynamics of these processes provides access to the features of that landscape.

Until recently, no technique existed that could deliver both single protein resolution and the temporal resolution required to quantify the dynamics of assembly. The advent of high-speed AFM (HS-AFM) with its sub-second temporal resolution and sub-molecular spatial resolution now makes such measurements feasible, stimulating an extensive research effort in this area[8-10] However, manually analyzing HS-AFM data, including identifying all the particles and tracking their trajectories to statistically describe the dynamics is not possible, due to the enormous volumes of data and presence of both instrumental noise and image artifacts associated with protein motion under an AFM tip. Similarly, even when the coordinates and orientation of the elementary building blocks are known as a function of time, extraction of physically relevant information is far from trivial. Indeed, whereas simulations yield the full information on positional and conjugate variables (e.g. forces) that allow full analysis of the physics of the system, the experimental observations generally yield only positional variables, necessitating development of analysis tools and workflows to deduce relevant physical laws and relationships. Correspondingly, further progress in understanding protein self-organization, as in many areas of observational sciences, necessitates the dual task of reliable materials-relevant feature extraction from the observational data and identification of correlational or generative models of the underlying physics from these descriptors.

Here we demonstrate a deep machine learning-based approach to extract the detailed dynamics of protein assembly from HS-AFM data. A deep neural network trained from a single labeled experimental image is used to identify the position and orientation of every protein over the course of assembly, despite the varying levels of noise and scan distortion. The output of the neural network is further analyzed in an unsupervised fashion to identify protein domains associated with distinct orientations. The full spatio-temporal diagram of domain trajectories is then reconstructed and the individual particle/domain trajectories are isolated and analyzed. Their relative pairwise energies are determined from the distribution of configurations. Finally, a Gaussian Mixture Model of particle morphologies combined with Markov analysis of the transition probabilities between states is used to get insight into the elementary mechanisms of particle behavior within the evolving microstructure.

Here we used HS-AFM to investigate *in-situ* the self-assembly of a *de novo* designed helical repeat protein DHR10-micaX (X=18), where X is the number of repeat subunits, on muscovite mica (*m*-mica) in 100 mM KCl solution. Inspired by ice-binding proteins, DHR10-mica18 has a designed *m*-mica binding surface on which 54 carboxylate residues geometrically match the Potassium ($K^+$) sublattice on the *m*-mica (001) plane (Fig. 1 (a)). In 100 mM KCl solution, DHR10-mica18 self-assembles on *m*-mica into a 2D liquid-crystal[3] consisting of discrete domains of 3 to 10 proteins that are co-aligned along one of the three closed-packed $K^+$ directions (Fig. 1 (b)). The domains remain dynamic, exhibiting fluctuations in size and orientation, as well as in the individual protein positions within the domains (Movie M1).

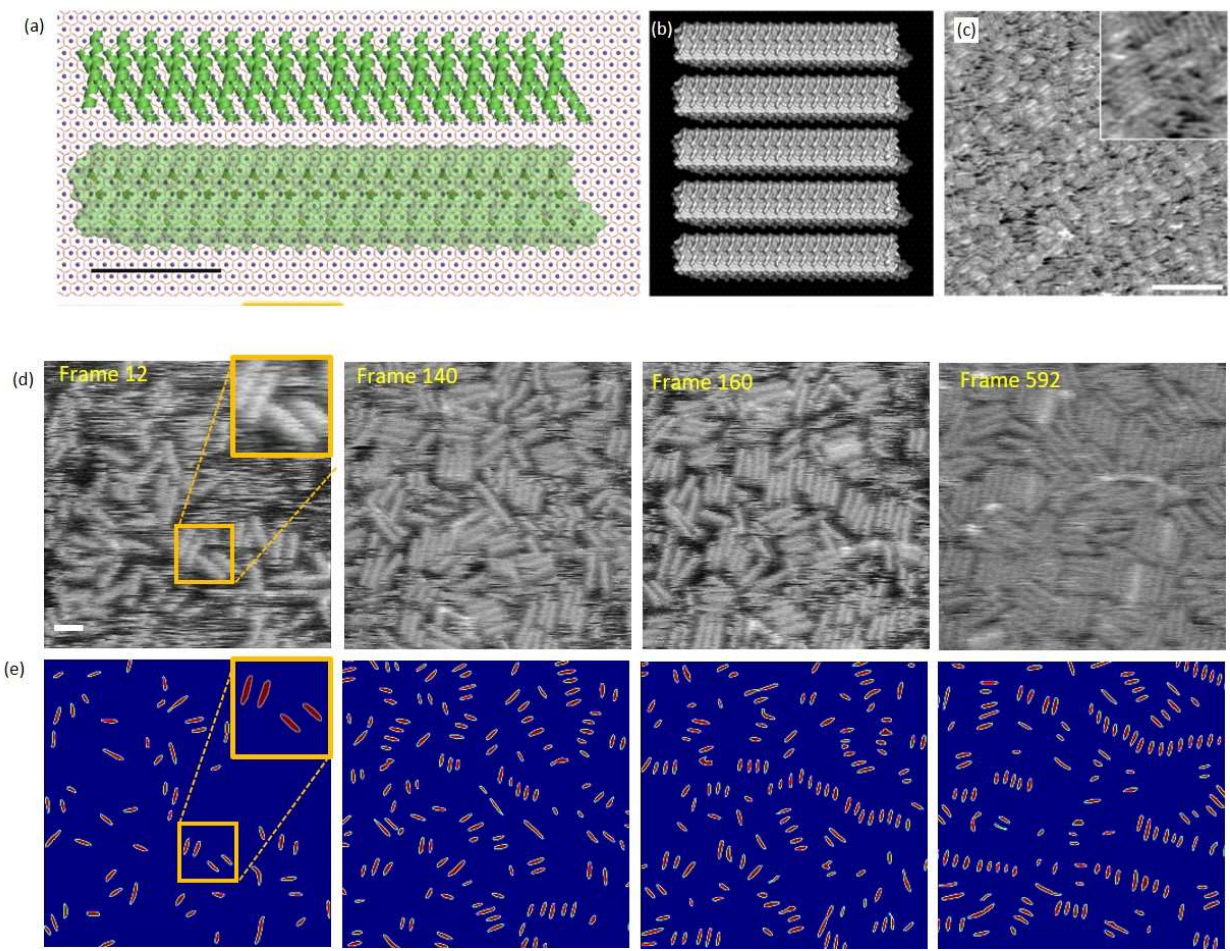

**Figure 1.** (**a**) Rosetta design models of two DHR10-mica18 proteins adsorbed on mica (001) surface via the K+ ion sublattice. Top: Cartoon view of DHR10-mica18 showing repeated alpha-helical secondary structure elements. Bottom: Semi-transparent view of the protein surface and lattice matched glutamate residues shown as sticks. Scale bar is 5 nm. (**b**) Model of five DHR10-mica18 proteins in co-aligned domain on $K^+$ ion sublattice of *m*-mica (001). Atoms are shaded according to their distance above the surface in a scale ranging from black (0 nm above surface) to white (2.5 nm above surface). (**c**) Conventional AFM image of DHR10-mica18 on *m*-mica (001) with 100 mM KCl. Scale bar is 100 nm (**d, e**) Individual time steps of protein crystallization (**d**) and corresponding deep-learning reconstructions (**e**) for 150 nm by 150 nm images collected at 0.38 Hz in a solution of 100 mM KCl and 0.05 uM DHR10-mica18. The scale bar is 15 nm. The solution contains 20 mM Tris buffer (pH=7).

Analysis of the dynamic AFM data on DHR10-mica18 assembly (Fig. 1) via machine learning represents a considerably more complex problem than previously reported analyses of scanning transmission electron microscopy (STEM) data.[11,12] This is because a STEM image can be well represented as a convolution of the delta-type functions representing locations of atomic nuclei with the microscope resolution function, which remains invariant along the image.[13] The variation of atomic weight alters the intensity of the signal, but not the resolution function and hence the feature shape. In comparison, an AFM image is formed by short-range interactions between the surface and the AFM tip,[14] and cannot be represented as a linear convolution. Furthermore, the particles have natural or apparent variability in shape and geometry. Similarly, the noise structure in STEM is generally uniform in the image plane, whereas the feedback operation in AFM results in proliferation of characteristic scan lines and streaks oriented in the fast scan axis direction. Correspondingly, for anisotropic objects as explored here, the noise effects can give rise to larger variability in the responses.

As a first step in the analysis, we adopted a deep learning (DL) model to convert noisy experimental data into the coordinates and orientations of surface particles using previously established approaches[15-17]. Specifically, we used a deep neural network from a U-net family of network architectures[18] to remove all the noise from experimental data and to classify each pixel in the input images as belonging to a specific class (in this case, class "particle" or class "background"). To generate the efficient training set for the problem, we utilized the fact that identification of individual particles improves with time as their motion slows and they form ordered domains, while the particles themselves do not change. Hence, the training set was prepared from a single image at the latest stage of particle ordering where the positions and orientation of all the particles are well-defined — usually the last frame of an AFM movie (Fig. 1 (d), frame 592). The image was labelled manually to yield a "ground truth". The training set was then generated via on-the-fly data augmentation by random cropping, adding Gaussian and Poisson noise, artificial scan lines and streaks, and zooming. In this manner, the prediction of a trained network should enable extraction of the position and orientation of each particle, because all possible combination of particle orientations, small distortions, anisotropic noise, and translation symmetries are now incorporated in the training set.

The trained neural network was applied to all the frames from the HS-AFM movie. The average decoding time for an image of 512x512 resolution was ~0.03 s with a NVIDIA's Tesla T4 GPU and ~1.5 s with on a standard desktop CPU. The output of the neural network is a map of "probability" (0 to 1) that each pixel from the input image belongs to a particle. We note that while in this case, there is only one class of particles, the same procedure can be extended to the classification of multiple particle types at the pixel level.

We then applied a binary thresholding ($th = 0.5$) to the output of the neural network and filtered out objects too small to be the targets of interest ($s < 50$ px$^2$). The remaining objects were fitted with an ellipse. The center of mass and angle (with respect to the vertical axes) of each ellipse were then stored for all the movie frames and used for further statistical analysis. We note that these frames represent the snapshots of the system at different times, representing the cross-sections across the individual particle trajectories. However, reconstruction of the individual trajectories (i.e. identification of same particle across the movie) especially in the presence of creation-annihilation processes represents a significantly more complex task that will be only partially addressed here. Correspondingly, first we explore the system evolution using the frame-by-frame analysis without trajectory extraction.

We first performed a statistical analysis of particle features extracted from the HS-AFM movie using the deep neural network in terms of classical descriptors, including pair distribution functions, 2D FFT, correlation functions, and angle distribution functions. In this analysis, the role of DL is to enhance the particle contrast, excluding the noise from the image. To quantify the average dynamic evolution of the system, we performed a multivariate statistical analysis of the average structural data. In this, the descriptor (e.g. time dependent 2D FFT or CF) was represented as a linear combination of the end members representing characteristic behaviors and the weights representing the evolution of the corresponding behavior with time. For example, for the 2D FFT we represent the full time-dependent Fourier transform as

$$FFT(k_x, k_y, t) = \sum_{i=1}^{N} A_i(k_x, k_y) w_i(t) \qquad (1)$$

where $FFT(k_x, k_y, t)$ is the full time dependent FFT transform for *t*-th image frame, $A_i(k_x, k_y)$ is the end member representing a specific behavior, and $w_i(t)$ is the time-dependent weight. The number of components $N$ is preselected before the decomposition and can be determined based on the quality of decomposition.

The decomposition method in Eq. (1) can be performed via a variety of unsupervised linear decomposition methods, such as principal component analysis (PCA), non-negative matrix factorization (NMF), gaussian mixture modelling, linear and non-linear autoencoders, etc. These methods differ in the specific constraints and weights, as well as the function optimized during the decomposition. Generally, the decomposition method should be chosen based on the physics of the signal and the extent of prior knowledge of the system, with the PCA being the simplest fully information theory-based method and NMF, Bayesian Linear Unmixing (BLU), etc. being progressively more complex methodologies that superimpose certain constraints on the endmembers and weights. For 2D FFT the natural condition is that the components $A_i(k_x, k_y)$ are non-negative, suggesting the use of NMF as the decomposition algorithm.

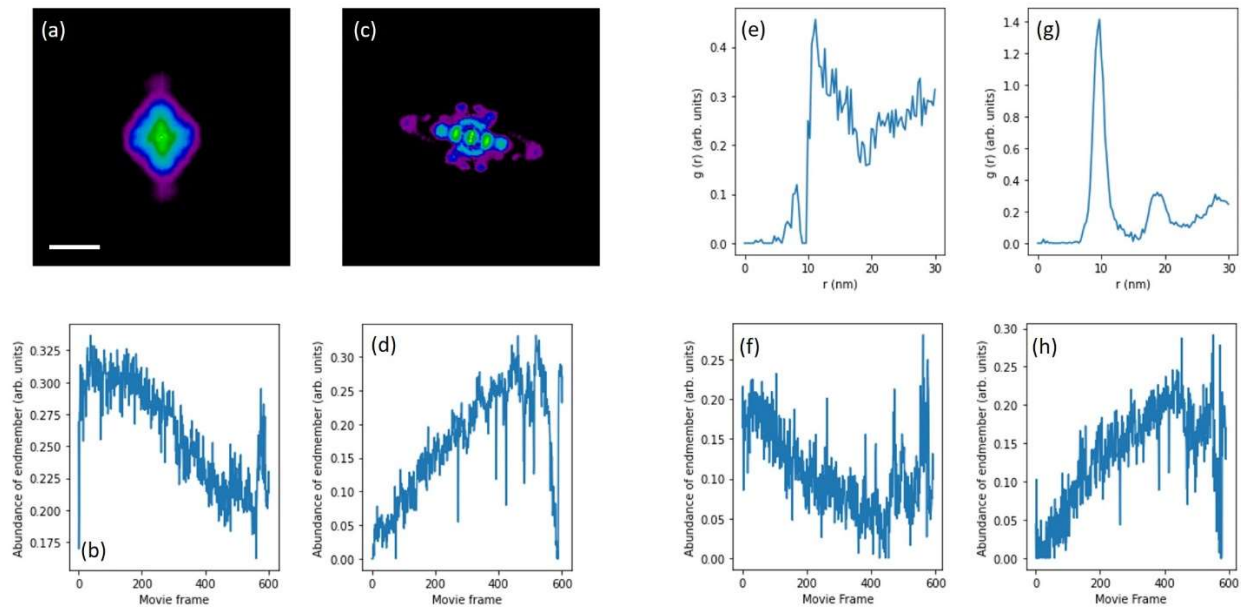

**Figure 2.** Multivariate analysis of the time dynamics in the nanoparticle system. NMF analyses of the 2D FFT including (a) first endmember (disordered phase) and (b) its temporal dynamics and (c) second endmember (ordered domains) and (d) its temporal dynamics. The scale is identical in (a) and (c), and the scale bar is 0.7 nm$^{-1}$. NMF analyses of the particle-particle correlation function (PP CF) including (e) first endmember (disordered domains) and (f) its temporal dynamics and (g) second endmember (ordered phase) and (h) its temporal dynamics. The decompositions for additional number of components are available in the accompanying notebook.

In the multivariate analysis of the 2D FFT transform of the system represented as a reconstructed image, the first NMF component shows the central peak only, with the intensity decreasing as a function of time (Fig. 2 (a,b)). The second component shows clear FFT peaks, with the overall intensity increasing as a function of time (Fig. 2 (c,d)). Note that the feature at frame ~580 corresponds to the loss of stability during AFM data acquisition. The decompositions into larger numbers of NMF components produces the more complex behaviors illustrating the gradual onset of ordering with time. Based on this analysis and comparison for different N, we limited the decomposition to two components only and associated the first one with the disordered and second with ordered phases of ellipsoids.

For the analysis of the temporal dynamics of the correlation function, we have similarly chosen NMF decomposition. Note that in this case we analyze the particle-particle correlation function defined on the particle center-to-center distances, rather than full original or reconstructed images. Shown in Fig. 2 (e,f) and (g,h) are the dynamics of the first and second NMF components, respectively. As with the 2D FFT, these clearly represent the dynamics of disordered and ordered components, respectively. Separation at larger numbers of components and use of PCA analyses allowed us to identify behaviors that can be associated with the changes in the first coordination sphere of the particles, but do not provide additional physics-based insights beyond that from the two component decomposition.

The analysis in Fig. 2 is based on the classical averaged measures of system behavior, roughly equivalent to the descriptors available from the scattering experiments. However, the real space images contain a significantly broader amount of information, potentially allowing much deeper insight into the system dynamics and evolution. The dominant feature of the particle organization dynamics is the formation of semi-ordered particle chains characteristic of liquid crystal behavior, with the particles having generally uniform orientation within the chain, but a significant degree of disorder manifest as random shifts along the normal directions. These particles exhibit very slow temporal dynamics. At the same time, a number of particles are trapped within regions with smaller particle density and exhibit considerably more dynamic behavior.

Here, we aim to get insight into these dynamics and explore the salient aspects of this behavior. To achieve this goal, we first calculated the average distance from each particle to its six nearest neighbors for all the movie frames and then plotted an "order parameter" defined as the number of particles with the same angle (± 5°) within the radius defined by the calculated average distance. We next performed an unsupervised classification of particles based on their orientations via a mean-shift clustering algorithm,[19] which is a non-parametric analysis technique for locating maxima of a density function given discrete data sampled from that function that allows different domains to be identified.

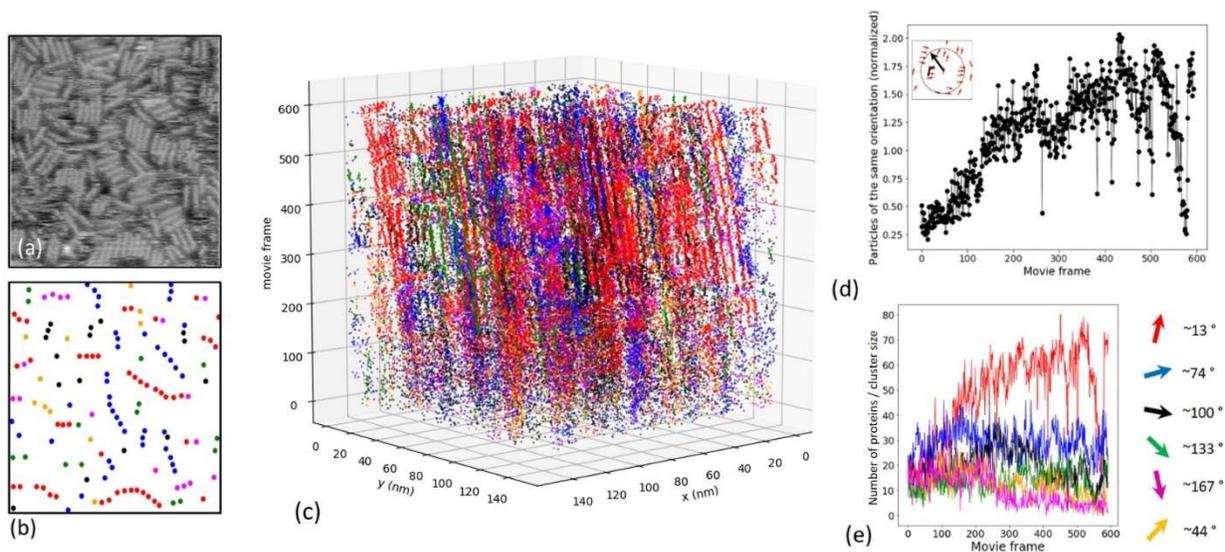

**Figure 3.** (a) Original image frame from later stages of system evolution and (b) center of mass positions of the individual particles colored based on similarity of orientation determined via mean-shift clustering. Note the clear identification of the individual chains and isolated particles in the disordered phase. (c) Full time-trajectory diagram of the system. (d) Evolution of the order

in the system defined as a number of particles with the same orientation (±5 deg) within the radius $r$, which is the average distance to the first six nearest neighbors calculated using all the movie frames (the number of particles within the radius is divided by the total number of detected particles in each frame). and (e) evolution cluster populations

This analysis is illustrated in Fig. 3 and associated movie M2. Shown in Fig. 3 (a) is one of the frames on the later stages of system evolution. The corresponding clusters are illustrated in Fig. 3 (b), with the points corresponding to the positions of the center of mass of the particle and colors corresponding to the cluster orientation index. Note that particles in similar states, including chain vs. isolated, and chain orientation are now clearly separated. The full temporal dynamics of the system can be represented as trajectories of individual particles in space-time, with the particle type indexed via color (Fig. 3 (c)). This approach hence enables separation of the jammed, static particles and relatively loose, moving particles. This in turn allows for analysis of the correlation functions, structure factors etc., for different particle groups (Fig. 3 (d, e)). Here, the evolution of the preponderant orientation is obvious.

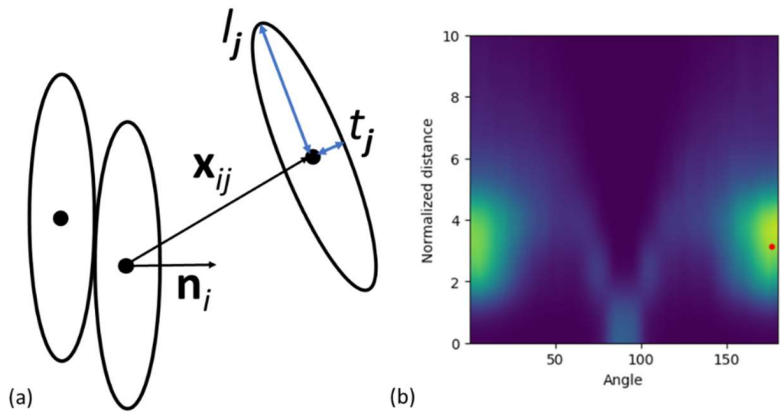

**Figure 4.** (a) Schematics of the particle geometry used for the analysis and (b) 2D histogram for normalized distance $d_{ij}$ and orientation $\theta_{ij}$ in degrees. Note that the maxima of the distributions for the assembled particles lie at orientation angles that differ from 0° or 180°, which corresponds to a slight tilt from co-alignment, and at normalized distances > 1, which corresponds to a particle-to-particle distance greater that the width of a single protein nanorod.

This analysis also allows us to gain insight into the interaction energy of the system. To achieve this goal, we parametrize the particle interactions via the normalized separation and misorientation, as shown in Fig. 4. Here, for each particle we define two closest neighbors, as dictated by the geometry of the problem. For each pair, we define the unit normal vector $\mathbf{n}_i$ for the $i$-th particle at the point closest to the geometric center of mass, and the radius vector $\mathbf{x}_{ij}$ to the $j$-th nearest neighbor ($j = 1,2$). The angle $\theta_{ij}$ between the two, $\cos(\theta_{ij}) = (\mathbf{n}_i \mathbf{x}_{ij})/\|\mathbf{x}_{ij}\|$ defines the misorientation between the particles, whereas the normalized distance, $d_{ij} = (\mathbf{n}_i \mathbf{x}_{ij})/(t_i + t_j)$ defines the contact conditions, where $t_i$ is the minor radius of the ellipse (i.e., half-width of the protein nanorod). By this definition, $d_{ij} = 1$ for particles in contact and $d_{ij} > 1$ for particles not in contact. For particles in contact, $\theta_{ij}$ defines the relative shift of the particles along the long axis. The probability distribution for the system $p(d, \theta)$ is shown in Fig. 4 (b). Here the two maxima corresponding to preferential particle positions are visible and correspond to a spacing that is greater than the width of a single protein nanorod and an orientation that is tilted slightly with respect to perfect coalignment.

The description of system evolution in Figs. 2 and 3 relies on human-defined physics-based descriptors. As such, the natural question is whether additional insight into the structure of the material can be derived from the machine learning based analysis of local geometries and structural building blocks. To achieve this goal, we perform the nearest neighborhood analysis for each particle, from which we construct a sub-image centered on the particle and oriented along the particle's long axis. This choice of the neighborhood as directly tied to the particle is fundamentally different from the classical sub-image selection used in sliding transform analysis of convolutional neural nets, and already relies on the known physics of the system. Also, in principle, the sub-images can be chosen both in the image frame (i.e. having the same orientation in image plane irrespective of particle rotation) and in the particle frame (oriented along the particle axis). The former approach will be more suited for the analysis of the particle-substrate interactions, whereas the second one is optimal for particle-particle interaction.

To analyze the local particle neighborhoods, we introduce the local shape analysis where we represent the collection of neighborhoods as a mixture of possible geometric shapes. Here, we use a Gaussian Mixture Model (GMM)[20-22] to represent the possible configurations. Specifically, we assume that there are $K$ Gaussian distributions parametrized by their mean $\mu_i$ and covariance $\Sigma_i$ and that each of $n$ extracted neighborhoods $R_i(x, y)$ is independently drawn from one of these distributions whose probability density function is given by:

$$p(\mathbf{R}_i) = \frac{1}{(2\pi)^{0.5n} |\Sigma_{k_i}|^{0.5}} \exp\left(-\frac{1}{2}(\mathbf{R}_i - \mu_k)^T \Sigma_{k_i}^{-1} (\mathbf{R}_i - \mu_k)\right).$$

The expected minimization algorithm is used to determine the parameters of the mixture with a pre-defined number of components. As in PCA and NMF, the determination of the optimal number of components in GMM represents a complex problem, which can be addressed based on the behavior of induvial components, quality of separation, and physics-based considerations. Here we note that, in the system with full rotational symmetry, rotationally invariant versions of the algorithm are necessary. However, in this case, the clear tendency of the system to form preferential orientations on the substrate allows us to use the classical GMM algorithm. The Bayesian Information Criterion (BIC)[23] shows a diffuse corner centered at ~10, suggesting that the classes in this system are poorly divisible, but the optimal cluster number is ~10.

This approach further allows us to explore the transformations between the different local configurations represented as GMM components. Here, we utilize the simple approach based on Markov transition probabilities for the trajectories of proteins labeled according to their GMM class. The protein coordinates were arranged into the trajectories using k-d-tree search method with a distance cut-off 4.7 nm. To discover the Markov matrix, we simply average the transition count matrix with its transpose.[24] Strictly speaking, Markov analysis can be applied only to a system in a quazi-stationary state, whereas Koopman formalism[25] should be applied to the non-ergodic system. Here, we assume that the Markov analysis is applicable during the latter stages of the AFM movie when the macroscopic order parameters are only weakly time-dependent.

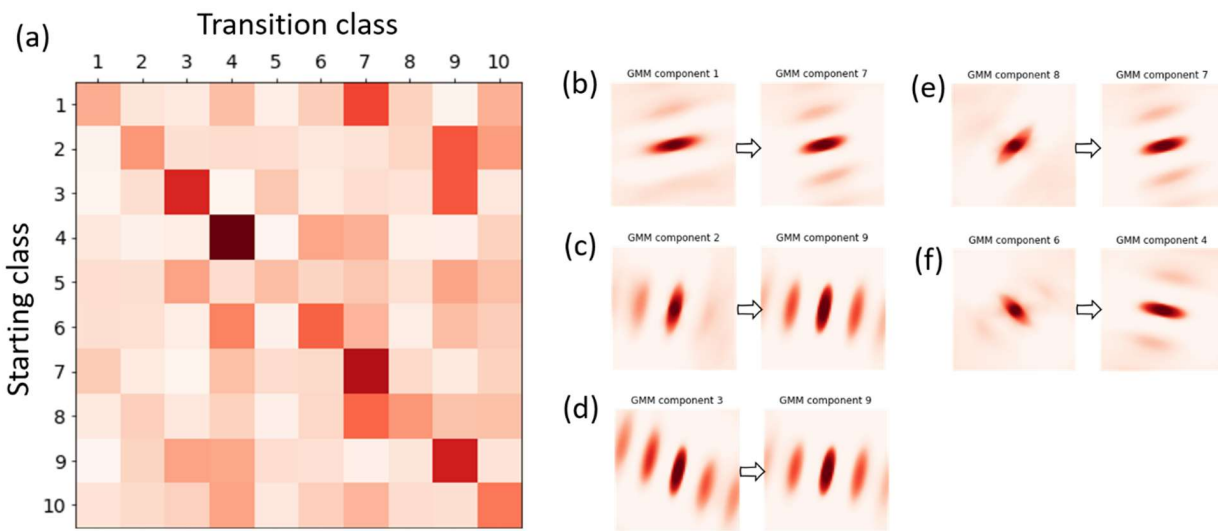

**Figure 5.** (a) The Markov transition matrix between different GMM components. The sum of elements in each row is 1. (b-f) The GMM states associated with highest transitions probabilities (excluding transitions to self) of 0.3 (b), 0.28 (c), 0.27 (d), 0.26 (e), 0.21 (f).

The transition matrix and the GMM states associated with highest transition probabilities (excluding transitions into self) are shown in Figure 5. Here, the transition in Fig. 5 (b) and 5(c) are associated with the formation of the ordered phase, while the transition in Fig 5 (d) shows change in the protein assembly lattice vector direction of the ordered phase. The transitions in Fig. 5 (e) and 5 (f) are associated with the switching of domain orientations (in the former, it is accompanied by formation of the ordered phase). This analysis hence allows disentangling transformation pathways within the ordered chains.

The results presented above show that a deep learning-based workflow enables the conversion of HS-AFM data into a map of the time dependent particle positions and orientations. This in turn allows reconstruction of classical two-point correlation-based descriptors, such as 2D FFT, and correlation and pair distribution functions. Principal component analysis and NMF, respectively can then be used to explore the time evolution from metastability towards equilibrium. Finally, machine learning-based feature extraction provides definition of particle neighborhoods and trajectories free of physics constraints. This approach then enables one to separate the possible classes of particle behavior, identify both the associated transition probabilities and slow modes and associated configurations, in turn allowing for systematic exploration and predictive modelling of the temporal dynamics of the system.

Our findings also have a number of implications for both analyzing HS-AFM data on protein self-assembly and developing a quantitative picture of the energy landscapes that govern the process. First, because a single frame contains hundreds to thousands of proteins, it provides an adequate data set from which to construct a robust neural network imbued with realistic noise and tip artifacts. Second, extracted particle descriptors provide both individual particle properties, such as positional and angular trajectories, as well as collective properties, like the distributions of configurations and their transitions as metastable orientations transition towards the stable orientations and the systems distributes itself into a time invariant set of domain sizes and orientations. Finally, once the system has reached equilibrium, the analysis provides the probabilities of finding specific configurations, as well as the probabilities of transitioning between individual domain configurations.

In the case of DHR10-mica18 on m-mica in 100 mM KCl solution, the results reveal an evolution from an initially disordered state in which coaligned domains are rare or non-existent to a final equilibrium state comprised of ordered domains containing up to 10 or more proteins. However, despite the continuing fluctuations of individual protein positions and orientations, as well as transitions between domain configurations, complete ordering is never achieved. This complex end states suggests competitive interactions between the proteins and with the substrate frustrate the development of nematic ordering expected for rod shaped particles with entropic interaction.[26] Future analysis of the full set of transitions should enable the extraction of relative free energies for each configuration and a reconstruction if the protein-protein and protein-substrate interaction potential that underlies that energy landscape. The codes developed in this work are publicly available in the form of interactive Jupyter notebooks at https://git.io/Jf1Wl .

**Acknowledgment:**

This research including AFM (S.Z., J.J.D.,), data analytics (X.L., O.D., S.V.K.) and protein synthesis (H.P., D.B.) are supported by the US Department of Energy, Office of Science, Office of Basic Energy Sciences, as part of the Energy Frontier Research Centers program: CSSAS–The Center for the Science of Synthesis Across Scales–under Award Number DE-SC0019288, located at University of Washington. The design of the proteins (H.P, D.B) was supported by the grant DE-SC0018940 funded by the U.S. Department of Energy, Office of Science. The machine learning was performed and partially supported (M.Z.) at the Oak Ridge National Laboratory's Center for Nanophase Materials Sciences (CNMS), a U.S. Department of Energy, Office of Science User Facility. High-speed AFM experiments were performed at the Department of Energy's Pacific Northwest National Laboratory (PNNL). PNNL is a multi-program national laboratory operated for Department of Energy by Battelle under Contract No. DE-AC05-76RL01830. The authors are grateful to Pratyush Tiwari (UMD) for valuable discussions.